\documentclass[prl,twocolumn,floatfix]{revtex4}

\usepackage{graphicx}
\usepackage{bm}

\newcommand{\be}{\begin{equation}}
\newcommand{\ee}{\end{equation}}
\newcommand{\ba}{\begin{eqnarray}}
\newcommand{\ea}{\end{eqnarray}}
\newcommand{\nn}{\nonumber \\}

\begin{document}

      \title{Does Adiabatic Quantum Optimization Fail for NP-complete Problems?}

      \author{Neil G.~Dickson}
      \author{M.~H.~S.~Amin}
      \affiliation{D-Wave Systems Inc., 100-4401 Still Creek Drive,
      Burnaby, B.C., V5C 6G9, Canada}

\date{\today}

\begin{abstract}
It has been recently argued that adiabatic quantum optimization would fail in solving
NP-complete problems because of the occurrence of exponentially small gaps due to crossing of local minima of the final Hamiltonian with its global minimum near the end of the adiabatic evolution. Using perturbation expansion, we analytically show that for the NP-hard problem of maximum independent set there always exist adiabatic paths along which no such crossings occur. Therefore, in order to prove that adiabatic quantum optimization fails for any NP-complete problem, one must prove that it is impossible to find any such path in polynomial time.
\end{abstract}
\maketitle

Adiabatic quantum optimization (AQO) was originally proposed \cite{Farhi01} as a possible means for solving NP-complete problems faster than classical computation. In AQO, the Hamiltonian of the system is evolved from an initial form, $H_B$, whose ground state defines the initial state of the system, to a final Hamiltonian $H_P$, whose ground state is the optimal solution to an optimization problem. To ensure a large amplitude of the ground state at the end of the evolution, the computation time $t_f$ should increase as $t_f \sim g^{-2}_{\rm min}$, where $g_{\rm min}$ is the minimum energy gap between the ground and first excited states during the evolution. The complexity of AQO is therefore determined by the scaling of $g_{\rm min}$ with the problem size.

Since its proposal, the complexity of AQO for solving NP-complete problems has been a
subject of controversy. Early works suggested possibility of polynomial scaling with
the size of the problem \cite{Farhi01}, but soon counterexamples were found
\cite{vanDam01,Znidaric06}. Later it was shown that the size of the gap in those
counterexamples can be increased significantly by changing the adiabatic path
\cite{Farhi02} or the initial Hamiltonian \cite{Farhi05}. However, until recently,
little insight existed on the physical process that can lead to exponentially small
gaps.

Using perturbation expansion, it was shown that a local minimum of $H_P$  crossing with the global minimum near the end of the evolution, sometimes called a first order quantum phase transition, can result in an extremely small $g_{\rm min}$, exponential in the Hamming distance between the two minima \cite{Amin09}. Using the same perturbation argument, Altshuler {\em et al.}~\cite{Altshuler} showed that for random exact cover instances, the probability of having such crossings increases with the system size and the crossing point moves toward the end of the evolution leading to an exponentially small gap. They, therefore, concluded that AQO would {\em fail} in solving random exact cover problems and possibly all NP-complete problems. Others also came to similar conclusions in different ways \cite{Young10,Jorg10}. Later, Knysh {\em et al.}~\cite{Knysh10} questioned the result of \cite{Altshuler} based on neglecting degeneracies of the minima and correlations between them. Moreover, the possibility of avoiding small gaps by changing adiabatic path was again pointed out by Farhi {\em et al.}~\cite{Farhi09}, and the fact that one problem can be mapped into many different Hamiltonians with different gap behavior was mentioned by Choi \cite{Choi10}. Those arguments, however, were based on numerical calculations for small problems, therefore inconclusive for large scales.

In this letter, we study the NP-hard \footnote{If any NP-hard problem can be solved in polynomial time, all NP-complete problems can.} maximum independent set (MIS) problem, into
which the exact cover problem can be mapped in polynomial steps \cite{Choi10}. Using perturbation expansion, we analytically show that: (i) For problems with non-degenerate local minima, or degenerate local minima distant from each other by more than 2 bit flips, it is trivial to choose an $H_P$ so that no crossing occurs between any local minimum (minima) and the global minimum. (ii) If $H_P$ has degenerate local minima, some exactly 2 bit flips apart, it is still always possible (although not as trivial) to avoid such crossings by changing $H_B$. Since there are infinite possibilities of choosing the total Hamiltonian, a valid proof of AQO failure must show that it is impossible to find an adiabatic path with no level crossing in polynomial time. Moreover, remaining in the ground state is not a necessary condition for solving NP-complete problems. As shown in Ref.~\cite{Zuckerman}, approximate solutions can also be used to solve NP-complete problems exactly in polynomial time. A proof of failure, therefore, must also show that no such approximate solutions can be obtained by AQO.

The MIS problem is that of finding a largest set $M$ of nodes in a
given graph $G$, such that there are no edges of $G$ between any
nodes in $M$.  If $n$ is the total number of nodes in $G$, the
problem of finding an MIS, $M$, can be expressed as minimizing a
cost function (energy) on $n$ binary variables, $x_i$, using
\cite{Choi08}
 \ba
 x_i &=& \left\{\begin{array}{cl}
  1 & \text{if $i\in M$} \\
  0 & \text{if $i\notin M$}
 \end{array} , \right. \nn
 E_P &=& - \sum_{i\in\, \text{nodes}} x_i + \sum_{(i,j)\in\, \text{edges}}
 c \, x_i x_j, \label{EP}
 \ea
with $c>1$. The last term in (\ref{EP}) is zero for every independent set of nodes,
because there is no edge between those nodes for which $x_i = 1$. Every dependent
set, on the other hand, gets a positive contribution from the sum for every pair of
adjacent nodes within it, thus resulting in a larger cost or energy. Therefore,
without the linear term, (\ref{EP}) would have a hugely degenerate global minimum
consisting of all independent sets. Notice that such a cost function (without the
linear term) has no local minima, because from every state it is always possible to
remove nodes (i.e., switch $x_i$ from 1 to 0) one by one to make all nonzero bilinear
terms vanish. Therefore, it is always possible to get from any state to the ground
state without ever increasing the energy. The role of the linear term in (\ref{EP})
is to assign different costs (energies) to different independent sets based on their
sizes so that the global minimum (or minima) of (\ref{EP}) becomes an MIS. There can
also be many local minima, which are {\em maximal} independent sets of the graph,
i.e., independent sets that cannot be made larger by adding nodes.

Our goal is to solve the MIS problem using AQO. We first represent every
node with a qubit by substituting $x_i {\rightarrow}
\frac{1}{2}(\sigma^z_i{+}1)$, where $\sigma^z_i$ is a Pauli matrix.
Equation (\ref{EP}), therefore, turns into a 2-local Hamiltonian
 \ba
 H_P = \sum_i h_i \sigma^z_i +  \sum_{i<j} J_{ij}
 \sigma^z_i\sigma^z_j, \label{HP}
 \ea
with $h_i = n_i c/4 - 1/2$, where $n_i$ is the number of edges connected to node $i$
(degree of $i$), and $J_{ij}=c/4$ ($=0$) whenever there is an (no) edge between $i$
and $j$. We introduce time evolution Hamiltonian
 \be
 H= H_P + \lambda H_B, \qquad H_B = - \sum_i \Delta_i \sigma^x_i,
 \label{H}
 \ee
where $\lambda$ changes from $\infty$ to 0.

Near the end of the evolution, where $\lambda {\ll}\, 1$, we can use perturbation
expansion in $\lambda$, taking $H_P$ as the unperturbed Hamiltonian, to calculate
eigenstates of $H$. Let $S$ be any maximal independent set (which could also be an MIS) of size $s$ and $|S^{(0)}\rangle$ be its corresponding state representing a local (or global) minimum of (\ref{HP}). From (\ref{EP}) we immediately find $E^{(0)}_S{=}-s$. Upon perturbation, the energy eigenvalue becomes $E_S(\lambda) = E_S^{(0)} + \lambda E_S^{(1)}  + \lambda^2 E_S^{(2)} + ...$.
The first order correction $E_S^{(1)} = \langle S^{(0)}| H_B | S^{(0)}\rangle = 0$,
because the operators $\sigma^x_i$ in $H_B$ flip one qubit and therefore, the overlap
of the wave functions vanishes. The lowest order non-zero term is therefore second
order, which as we shall see is the dominant one.

Now suppose that $|M^{(0)}\rangle$ is the global minimum and $|M'^{(0)}\rangle$ is a
local minimum of (\ref{HP}), each representing a maximal independent set ($M$ and
$M'$) of size $m$ and $m'$ ($<m$) respectively. Let $\delta E(\lambda) \equiv
E_{M'}(\lambda){-}E_M(\lambda) = \delta E^{(0)} {+} \lambda^2 \delta E^{(2)} {+}
O(\lambda^4)$ denote the energy separation between the two states. To zeroth order,
$\delta E^{(0)} {=} m{-}m'{>} 0$, as expected since $M$ is the MIS. If at some
$\lambda$ within the convergence radius of the perturbation, $\delta E(\lambda)<0$
for a finite order of perturbation \footnote{In principle, infinite order
perturbation beaks down at the avoided crossing. However, finite order perturbation
may still provide a reliable asymptotic expansion for the eigenvalues.}, this means
that at some point $\lambda=\lambda^*$ the two levels should cross. This can lead to
an extremely small minimum gap at the (anti)crossing point, which is exponentially
dependent on the Hamming distance between the two minima \cite{Amin09}.

We first consider the simplest case with no degeneracy and a uniform
transverse field, $\Delta_i=1$. The second order correction to the
energy is (with $S=M,M'$)
 \ba
 E_S^{(2)} = \sum_{k \ne S} {\langle S^{(0)}| H_B |k^{(0)}
 \rangle \langle k^{(0)}| H_B | S^{(0)} \rangle \over
 E_S^{(0)} - E_k^{(0)}}. \label{E2}
 \ea
Since $H_B$ involves all single bit flip operators $\sigma^x_i$, the
sum is nonzero only for states $|k^{(0)}\rangle$ that are 1 bit flip
away from $|M^{(0)} \rangle$:
 \ba
 E_S^{(2)} = -\sum_i{1\over B_i}, \qquad
 B_i = \left\{\begin{array}{cl}
  1 & \text{ if $i\in S$} \\
  cd_i-1 & \text{ if $i\notin S$}
 \end{array}\right. ,
 \ea
where $B_i$ is the energy cost of flipping qubit $i$ from state $|S^{(0)}\rangle$,
and $d_i$ is the number of edges between $i$ and nodes in $S$. Notice that $d_i\ge
2$, since $d_i=0$ implies that node $i$ is not connected to any nodes in $S$, i.e.,
$S$ is not maximal, and $d_i=1$ implies that there is a degenerate state 2 bit flips
away. Up to second order in $\lambda$, we find
 \ba
 E_S(\lambda) = -(1+\lambda^2)s - \lambda^2 \sum_{i\notin S} {1 \over cd_i-1}.
 \label{EMl}
 \ea
Since $s$ is largest for the MIS, the first term in (\ref{EMl}) is minimum for the
global minimum as desired. The sum, however, depends on $d_i$ and could be larger for
a local minimum than the global minimum. As a result, at large enough $\lambda$, it
can cause the two energy levels to cross. However, since $c$ can be chosen to be
arbitrarily large, it is possible to reduce the effect of the sum and therefore
eliminate the anticrossing by increasing $c$. Physically, increasing $c$ will push
low lying excited states (specifically dependent sets) adjacent to the global and
local minima upward while keeping the minima energies fixed. This would reduce the
relative magnitude of $E_S^{(2)}$ in (\ref{E2}), as it inversely depends on the
excited state energies. For example, by choosing $c=n$, we get
 \ba
 \delta E(\lambda) > (1{+}\lambda^2)(m{-}m')
 - \lambda^2 \frac{n{-}m'}{2n{-}1} > 1{+}{\lambda^2 \over 2},
 \ea
because $m{-}m'\ge 1$, $m'\ge 1$, $d^\prime_i\ge 2$, and the number of nodes
not in $M^\prime$ is $n{-}m'$. This is strictly positive, therefore
no anticrossing occurs between those minima.

The above argument holds beyond second order perturbation. Assuming $c{=}\,n$ and
that a minimum $S$ is sufficiently isolated from other minima so that up to $q$ bit
flips the energy of the state always increases, the $q$th order perturbation
correction to the energy is
 \ba
  E_S^{(q)}= \frac{(-1)^{q/2}(q-2)!}{(q/2-1)!(q/2)!} s + O(1).
  \label{EMq}
 \ea
The radius of convergence deduced from (\ref{EMq}) is $\lambda_c{=}\frac{1}{2}$, and
the contribution from each successive term is monotone decreasing within $\lambda_c$,
so the second order term remains dominant. Moreover, neglecting the $O(1)$ part of
(\ref{EMq}) compared to the first $\Theta(s)$ term, we find that up to the $q$th
order perturbation, where validity of (\ref{EMq}) stands, we can write
$E_{M'}(\lambda) = (m'/m) E_M(\lambda)$, therefore $\delta E(\lambda) =
[(m'{-}m)/m]E_M(\lambda)>0$, which follows from $m>m'$ and $E_M(\lambda)<0$. This
means that up to the $q$th order perturbation, the levels do not cross.

The price to pay for eliminating the level crossings is to increase the coupling
constant $c$ linearly with the size of the problem. Alternatively, one can keep $c$
constant, but divide the linear terms in (\ref{EP}) by $n$ to achieve the same goal,
as was done in \cite{Choi10} (although the local minima in \cite{Choi10} are
degenerate). This leads to smaller energy steps in the spectrum of $H_P$ and
therefore higher required precision. In both cases, the scaling of the coupling
constant or precision with $n$ is polynomial (linear), whereas the gain in
eliminating the crossings could be exponential.

One may argue that by increasing $c$ with $n$, or equivalently dividing the linear
terms in (\ref{EP}) by $n$, we reach a regime in which the bilinear terms in
(\ref{EP}) become dominant and determine the dynamics of the system. Thus, although
there is no level crossing for $\lambda {\ll} 1$ considered above, there could be one
for $1 {\ll} \lambda {\ll} n$. In this region, one can neglect the linear terms in
(\ref{EP}) and repeat the above perturbative argument keeping only the bilinear
terms. However, as mentioned earlier, the Hamiltonian without the linear terms has no
local minima, but only a hugely degenerate global minimum consisting of all
independent sets. Such a Hamiltonian, therefore, cannot produce a crossing in the way
discussed above.

So far, we have only considered non-degenerate states, which is indeed the level of
discussion in Refs.~\cite{Altshuler,Farhi09}. We now take a step further and discuss
cases with degenerate minima. We first have to generalize (\ref{E2}) to include
degeneracies. Suppose there are $K$ maximal (or maximum) independent sets $S_k$ of
size $s$, with $k=1, ..., K$. States $|S_k^{(0)}\rangle$ and also every superposition
of them are therefore degenerate eigenstates of $H_P$. Perturbation in $\lambda$
removes this degeneracy. Let $|S^{(0)} \rangle {=} \sum_k C_k |S_k^{(0)} \rangle$
represent the lowest energy superposition after the degeneracy is lifted. With the
positive sign of $\Delta_i$, $C_k$ will all be positive real numbers with the
constraint: $\sum_{k}C_k^2 = 1$. The first order correction is zero because all $S_k$
are the same size ($s$) and therefore one cannot get from one minimum to another by a
single bit flip (adding or removing a single node). The second order correction is
 \ba
 E_S^{(2)} &=& -\sum'_{(k,k'),(i,j)}
 {\Delta_i\Delta_j C_kC_{k'} \over  B_{k,i}},
 \label{E2deg}
 \ea
where $B_{k,i}$ is the cost of flipping qubit $i$ from state $|S_k^{(0)}\rangle$, and
the prime sign on the sum means that the sum is over all the paths from
$|S_k^{(0)}\rangle$ to $|S_{k'}^{(0)}\rangle$ with two bit flips by first flipping
qubit $i$ and then qubit $j$.

If there are no two minima $|S_k^{(0)}\rangle$ and $|S_{k'}^{(0)}\rangle$ that are
exactly 2 bit flips distant from each other, (\ref{E2deg}) becomes similar to the
non-degenerate equation (\ref{E2}). In that case, the argument deduced from
(\ref{E2}) holds, i.e., level crossings can be eliminated by increasing $c$ linearly
with $n$. The exceptions, therefore, are cases with minima 2 bit flips apart from
each other. The worst cases would have only one global minimum but numerous local
minima 2 bit flips apart.  The negative contribution to $E_{M'}^{(2)}$ can then
become large enough to bring down the total energy of the local minima below that of
the global minimum at a $\lambda < \lambda_c$. This is the case we shall consider
now.

To eliminate the above crossing, we need
to make the energy difference $\delta E(\lambda) = m{-}m'
{+} \lambda^2 \delta E^{(2)}$ strictly positive. This is achieved if
 \ba
 \delta E^{(2)} = \sum_{i} {\Delta_i^2 \over B_i}
 - \sum'_{(k,k'),(i,j)} {\Delta_i\Delta_j C_kC_{k'} \over  B'_{k,i}}
 > 0.
 \ea
Here, $B_i$ ($B'_{k,i}$) is the cost of flipping qubit $i$ in the
global minimum $M$ (local minimum $M'_k$). A sufficient condition,
which is independent of $C_k$, is
 \be
 \sum_{i} {\Delta_i^2 \over B_i} -
 \max_k \sum'_{k',(i,j)} {\Delta_i\Delta_j \over B'_{k,i}} > 0.
 \label{cond}
 \ee
This can be proved using the fact that for every non-negative, symmetric matrix $A$
operating on unit vectors $v$, we have: $\max_v v^TAv \le \max_k \sum_{k'} A_{kk'}$.
For simplicity, here we focus on the large $c$ regime (e.g., $c{=}\,n$) for which the
states that violate edges can be neglected. All calculations can be generalized to
the small $c$ regime, but the equations become more complicated. Condition
(\ref{cond}) becomes ${\cal F}(\{\Delta_i\}) > 0$, where
 \be
 {\cal F}(\{\Delta_i\}) \equiv \sum_{i\in M} \Delta_i^2 -
 \max_k \sum_{i \in M'_k}\left[\Delta_i^2
 + \sum'_{k',j} \Delta_i\Delta_j \right]. \label{F}
 \ee
Since there is freedom in choosing values of $\Delta_i$, one can choose them such
that  ${\cal F}(\{\Delta_i\}) > 0$. A successful assignment makes the first term in
(\ref{F}) large and/or the second term small so that the result becomes positive. A
trivial choice is $\Delta_{i\in M} = \alpha$ and $\Delta_{i\notin M} = 1$. Let
$p_k=|M\bigcap M'_k|$. Since $M'_k$ is a local minimum, we have $p_k\le m-2$. Also,
there are at most $n-m'$ local minima all 2 bit flips away from $M'_k$. Therefore,
 \ba
 {\cal F}&\ge&\alpha^2 m - \max_k \left( \alpha^2 p_k + m'-p_k + \alpha(n-m') \right) \nn
 &\ge& 2\alpha^2 - \alpha(n-m') - m'+1,
 \ea
which is positive if
 \ba
 \alpha > \frac{n-m'+\sqrt{(n-m')^2 + 8(m'-1)}}{4}
 \approx \frac{n-m'}{2}.
 \ea
Such an assignment, however, is not very useful, since it assumes knowing the
solution to the problem. Nevertheless, it proves the existence of at least one
assignment of $\Delta_i$ for which no crossing occurs.

We now consider another assignment which does not assume the solution. Let $M' {=}
\bigcup_k M'_k$ and $p=|M\bigcap M'|$. We assign $\Delta_{i \in M'} = \beta$ and
$\Delta_{i \notin M'} = 1$. The last term in (\ref{F}) is multiplied by $\beta^2$ and
the first term becomes $m-p+\beta^2p$. We find ${\cal F} \ge m-p+\beta^2p - \beta^2[m'+n-m'].$ In order for ${\cal F} > 0$, it is sufficient to have
 \be
 \beta < \sqrt{(m-p)/(n-p)}.
 \ee
Therefore, as long as $m>p$, that is as long as there is at least one node in the
global minimum that is not in any of the local minima, there exists an assignment of
$\Delta_i$ for which no crossing happens during the adiabatic evolution. As before,
the scaling of $\beta$ with the size of the system is only polynomial.

This assignment does not require knowledge of the global minimum (solution), but only
knowledge of the local minima, which can be obtained by running the
evolution multiple times with $\Delta_i=1$. If there is a crossing
between the global minimum and the state comprising the local minima, every time the
system does not reach the global minimum, it falls into one of the local minima.
Moreover, other degenerate local minima in the neighborhood of the one reached can be
obtained in polynomial time using local search. The more information is obtained from
such suboptimal evolutions, the more one can adjust $\Delta_i$ to avoid the crossing.

The above assignment may not eliminate all such crossings if $m=p$. However, there
are an infinite number of ways to define the input parameters. For example, one can choose nonuniform $\Delta_i$ within $M'$ or nonuniform $J_{ij}$.
Combining these and other ideas may also give a Hamiltonian that satisfies
(\ref{cond}). There are therefore infinitely many possibilities to define a
Hamiltonian for solving an NP-hard problem instance, many of which may not have level
crossings. An iterative numerical method that follows from the above ideas has proven
to be successful in eliminating crossings in extremely difficult instances with
highly degenerate local minima, even where $m=p$ \cite{Dickson}.  Other methods have also been proposed for finding an optimal path \cite{Rezakhani}.

It is important to note that we are not trying to prove that all level crossings
between a global minimum and local minima can be eliminated in polynomial time.
Neither are we claiming that if they are eliminated, the MIS problem can be solved in
polynomial time. Even if all level crossings are eliminated, the scaling of the
minimum gap in the rest of the spectrum is still unknown. What we are stating here is
that there always exist paths along which no crossing occurs, at least up to second
order perturbation. Since MIS is NP-hard, any NP problem can be polynomially mapped
onto it. Therefore, a valid proof that any NP-complete problem cannot be solved using
AQO because of level crossings must prove that for the problem mapped onto MIS, it is
impossible to find an assignment of parameters for which there is no level crossing.
Further, due to the NP-hardness of {\em approximating} solutions to MIS
\cite{Zuckerman}, even if there are multiple crossings, AQO may produce sufficient
solutions to solve NP-complete problems.

In conclusion, using perturbation expansion, we have shown that for the NP-hard
problem of MIS, it is always possible to write down a Hamiltonian for which during
the adiabatic evolution no crossing occurs between a global minimum and any of the
local minima. If there is no degeneracy in the local minima, or if
there are degenerate local minima but no pair of them is exactly 2 bit flips apart,
such a Hamiltonian can be trivially obtained by increasing the coupling coefficient
between the qubits linearly with the size of the problem. In cases with local minima
exactly 2 bit flips away from each other, one can use the freedom of choosing the
initial Hamiltonian to avoid level crossings. In the latter case, finding an
assignment for tunneling amplitudes $\Delta_i$ may or may not be nontrivial. However,
we have shown that such an assignment always exists. In general, there are infinite
ways of defining the Hamiltonian, including those where many approximate solutions
suffice, therefore it seems infeasible to prove that no successful Hamiltonian can be
obtained in polynomial time.

We thank B.~Altshuler, P.~Chavez, V.~Choi, S.~Gildert, F.~Hamze, K.~Karimi, R.~Raussendorf, C.~Rich, G.~Rose, and A.P.~Young for useful discussions.

\end{document}